\documentstyle[prl,twocolumn,aps]{revtex}

\begin{document}
\draft
\title{Objectivism and Irreversibility in Quantum Mechanics}
\author{ Takuya Okabe }

\address
{
Faculty of Engineering, Shizuoka University, 
Hamamatsu 432-8561, Japan
}

\date{
\today
}
\maketitle

\begin{abstract}
A hypothetical formulation of quantum mechanics 
is presented so as to reconcile it with macro-realism.
On the analogy drawn from thermodynamics,
an objective description of
wave packet reduction is postulated, in which
a characteristic energy scale
and a time scale are introduced to separate the 
quantum and classical conceptions.
\end{abstract}

\section{INTRODUCTION}
Despite the indisputable practical success of quantum mechanics (QM), 
conceptual and philosophical difficulties are still left
behind.\cite{rf:bd,rf:lg}
In fact, we are so accustomed to the classical notion 
in real life that
we cannot even imagine what a superposition of macroscopically distinct
states, as stipulated by QM as a possibility, really looks like. 
%
Here we aim to 
present a hypothetical formulation
to fill in the conceptual gap between 
classical and quantum mechanics.
First of all,
let us state our standpoint 
to tackle this long-standing problem by three steps, 
on each of which respectively one will find many proponents as well as
opponents.

Firstly, we regard a wave function as elements of reality, 
characterizing an individual physical system, not an ensemble.
We believe that the quantum formalism is something more than merely a
set of observational predictive rules.
We pay attention to time evolution of 
individual systems
and 
assume that the wave function gives the fullest description of 
a quantum state.
In this interpretation, 
wave function collapse is ultimately unavoidable,
and we shall regard the collapse as a {real and physical} process.
At this point, we deviate not only from the Copenhagen interpretation,
but also 
from the Everett relative-state interpretation.

Secondly, as there must be no privileged observer, 
we assume that
the wave packet reduction is a {spontaneous} process.  
In this respect, there now exists a class of notable theories
stemming from the original work of Ghirardi, Rimini and Weber
(GRW),\cite{rf:grw,rf:grw2}
in which the time evolution equation of standard quantum theory 
is elaborately modified by introducing nonlinear and stochastic
elements.
In this approach, the classical behavior of macroscopic systems
as well as the quantum properties of microscopic systems are 
derived altogether from a unified dynamics.
Though not related directly,
there are other works similar in spirit
but from a different perspective, viz.,
quantum gravity.\cite{rf:qg,rf:rp}
These theories predict
testable consequences against standard quantum theory.
Most notably, one of their striking outcomes is 
the violation of energy conservation.\cite{rf:grw,rf:qg,rf:ps}
Somehow they introduce a characteristic scale 
to describe a crossover from
the microscopic quantum regime to the ordinary macroscopic regime.
The latter emerges from the former by state vector reduction.
The nonlinear theories are soundly motivated 
by the crucial point of QM that
the essential problem
is indeed a consequence of 
the general and unavoidable fact that 
the state space as well as 
the evolution equation thereof are linear. 
In fact, 
it is often remarked\cite{rf:bd2} that there is no way out
to render the quantum formalism ontologically interpretable
but either to 
alter it in more or less ad hoc ways,
by plugging in nonlinear terms 
in the Schr\"odinger equation,  
or to assume explicitly hidden variables.
Nevertheless,
this remark is implicitly based on 
an expectation
that time evolution of a `state', 
whatever it means, must be governed by a {\it unified}
dynamics.  It is indeed convincing, but this is the point where
we break
away from various types of nonlinear theories.
Such unified theories could afford to make quantitative, but not
qualitative, difference between two categories
of natural phenomena.

As the third point, discarding the assumption of
a unified dynamics or a single prescription,
we propose to promote
the wave function collapse to the status of the elementary process,
ranked along with the unitary linear time evolution.
Our approach consists in accepting 
from the outset that 
wave functions can develop temporally via two distinct ways.
We do not modify the unitary evolution at all,
but supplement it with a subsidiary condition
for the collapse to intervene occasionally.
The condition will be expressed by an inequality.
We come to adopt this strategy
in consideration of 
the status quo that
any such attempts to {\it derive}, or {\it explain} the collapse process
from a more fundamental level 
must be faced with more or less 
conceptual or mathematical difficulties at that
level.\cite{rf:bg,rf:ggp}
To get around them virtually,
we shall take the less ambitious attitude in a sense, 
i.e., to renounce the attempts altogether at the outset.
Indeed it is as old as quantum mechanics itself
to postulate two fundamentally distinct laws of evolution,
but in this context the second type 
has almost always been attributed
to the act of consciousness.
Divorced from such subjective approaches,
we intend to formalize objectively 
the reduction process on the premise 
that the reduction is fully characterized by the input and the output
states, without delving into the mechanism in between.

In our approach, 
quantum states are almost always governed by the linear dynamics,
except when the probabilistic collapses happen.
The collapse mechanism operates only under a specific condition,
which is to be identified.
For our purpose, 
a remark due to von Neumann is especially noteworthy,\cite{rf:jvn}
that the two kinds of processes can be distinguished unambiguously
by the concept of statistical entropy.
Nevertheless, it is easily conceivable
that von Neumann's entropy criterion is
not sufficient.
To find an answer to 
the problem of `measurement', 
we need to identify not only the boundary between 
deterministic and stochastic, or reversible and irreversible,
but also that between `microscopic' and `macroscopic', 
hopefully simultaneously.
The latter boundary
between system and observer
is often cited as a principal cause of debate, 
which originates from the very fact 
that there is nothing in standard quantum theory
to fix such a borderline.

Bearing in mind the conceptual achievements of the collapse
theories,\cite{rf:grw,rf:qg,rf:pp,rf:gpr} and
in the full conviction that
there {\it must} be the definite boundary somewhere 
in the middle of the `micro' and `macro',
we aim to remedy the above drawback by looking for the objective
criterion, that, if successful, would replace
some perceiving subject 
which von Neumann, Wigner and others have had recourse to.
To fix the borderline,
we will bring in some parameters as naturally
as possible,  
without introducing any other unobservable machinery agent.
In a sense,
our attempt may be viewed as a step toward objectification of the wave
packet reduction postulate,
by which to judge if a given linear superposition is stable or unstable.
We see what emerges from the synthesis to appear thereafter.
It is anticipated that any unstable superposition 
is doomed to collapse at random,
in accordance with the probability principle of QM.
As a fundamental rule, 
the criterion must be simple and appealing.
Moreover, as a stringent prerequisite,
the new formalism must not spoil
the statistical predictions of QM for microscopic systems,
as they have been overwhelmingly confirmed without any doubt.
To put it concretely,
we are concerned about
whether or not a given state collapses, or
whether the system is `observed' in a given situation. 
For example, an electron is `observed' 
when injected in 
a cloud chamber, but not
when bound 
in the ground state of 
a hydrogen atom.
When `observed', we need not only
reproduce the predicted results as expected,
but should abstract the presumed condition that is met there.

In search of a satisfactory formulation of QM,
we find it instructive to cite the two guidelines according to
J.~S.~Bell.\cite{rf:bell}\footnote{
The quoted sentences,
published prior to the GRW theory,
are meant to introduce 
the pilot wave interpretation of de Broglie and Bohm.
}
``The first is that it should be possible to formulate them for small
systems.'' 
(Otherwise, it is likely that `laws of large numbers' are being invoked
at a fundamental, so that the theory is fundamentally approximate.)
``The second, related, point is that 
the concepts of `measurement', or `observation', or `experiment',
should not appear at a fundamental level.''
(Because
these concepts appear to be too vague to appear at the base of a
potentially exact theory.)

It is the core of the paper
to postulate the criterion in Section \ref{trigger}.
Before that, 
in order to see how the criterion
serves its purpose,
we provide a framework 
to discuss wave packet reduction
in Section \ref{preferredbasis}.
Section \ref{discussion} contains some applications of the theory.
A brief summary is presented in Section \ref{summary}.

\section{Preferred basis problem}
\label{preferredbasis}

States resulting from wave packet reduction are not arbitrary.
Before discussing the condition for the reduction,
we have to specify on what basis a given wave function is projected 
by reduction.
This is called the preferred basis problem.\cite{rf:pp}
The GRW theory pays special attention to a position basis, 
onto which a wave function is reduced to 
realize a localized state in the real space representation.
At a glance over the general structure of the transformation theory
of QM, however,
there seems no special reason to prefer the position
basis,
since we can think of many other types of linear superpositions
which are equally as clumsy 
but not necessarily consisting of spatially far-off separated parts.
In fact, 
a more general basis 
is apparently required in practice
to accommodate with
various kinds of macroscopic quantum phenomena
of current experimental interests,\cite{rf:mqp}
although it may well be argued that 
the measurement problem should in any case 
boil down to the projection onto the position basis 
of an observer or apparatus.\cite{rf:abgg}
In this section, 
we provide a framework within which to discuss the wave
packet reduction phenomena on a general basis.
In that, we aim to substantiate an insight that
wave packet reduction is a process to break off
weak coupling correlation developed in a wave function.

A normalized solution $\Psi$ of the wave equation,
\[
 i\hbar \frac{\partial \Psi}{\partial t} = H \Psi, 
\]
is expanded in terms of a complete set of ortho-normalized functions 
$\Phi_n$ as
\[
 \Psi=\sum_n c_n \Phi_n. 
\]
According to QM, 
the probability $w_n$ for the initial state $\Psi$
to be `observed' in the state $\Phi_n$
is given by 
\[
w_n= |c_n|^2=|\langle \Phi_n|\Psi\rangle|^2. 
\]
For instance,
the state $\Phi_n$ represents a product state of
the $n$-th `reading' of an apparatus 
and the corresponding state of a system.
As an outcome of reduction,
$\Phi_n$ will possess
some `classical' properties of the apparatus.
To represent the `classical' basis,
let us introduce the Hamiltonian $H_0$
which is diagonal in the representation $\Phi_n$,
\[
 H_0 \Phi_n = E_{n} \Phi_n. 
\]
Since $\Phi_n$ need not be an eigenstate of the true Hamiltonian $H$
of the whole system, 
we will generally find $H=H_0+H_1$ and $[H_0,H_1]\ne 0$.
In practice, the difference $H_1=H-H_0$ may be formally regarded as a
negligible 
quantum-mechanical perturbation to $H_0$, 
so that the `macro' state
$\Phi_n$ would be only quasi-stationary.

To put it differently,
let us assume an eigenstate $\Phi_i$ of $H_0$
to represent a `macroscopic' state.
In the presence of an ever-present microscopic off-diagonal
perturbation $H_1$, and in the absence of any `observer',
it will be developed into a grotesque
linear superposition by
the causal time evolution,
\begin{equation}
\Phi_i\rightarrow 
 \Psi=\sum_n c_n \Phi_n.
\label{(1)}
\end{equation}
However small $H_1$ may be,
this is generally an ultimately inevitable consequence,
in principle.
Therefore, we have somehow recourse to the reduction
process projecting $\Psi$ back again onto one of the `macroscopic' states,
$\Phi_m$.
In terms of the wave function,
\begin{equation}
\Psi=\sum_n c_n \Phi_n
\rightarrow  
\Phi_m,
\label{(2)}
\end{equation}
or, in terms of the density matrix
\begin{equation}
 \hat{\rho}=\sum_{m,n}c_m^*c_n|\Phi_n\rangle\langle\Phi_m|
\rightarrow  
\sum_{n} w_n |\Phi_n\rangle\langle\Phi_n|.
\end{equation}

The reduction process is characterized not only by the resulting set of
states, but also by the loss of phase correlation 
possessed by the initial linear superposition $\Psi$.
We suppose
that the both aspects are built into the Hamiltonian
as
$H=H_0+H_1$ in the way that
$H_0$ provides a basis set on which
the initial entangled configuration is projected,
while $H_1$ characterizes the correlation to be lost in the reduction.
In effect, all the situations 
encountered in `measurement' seem to
have the Hamiltonian of this structure.
In practice,
the off-diagonal matrix elements of $H_1$ may quantitatively
depend on the choice of the basis defined with $H_0$.
By way of illustration,
we will later regard the kinetic energy of 
a massive particle as $H_1$ in order to realize 
a spatially localized wave packet.
Then the matrix elements will depend on
the spatial width of the resulting localized state.

The above discussion is 
aimed at providing a framework in which to discuss 
the projection postulate objectively
in the following sections.
In any case of our concern,
the decomposition of the full Hamiltonian $H$
into an `independent-particle' basis and `(scattering) interaction'
thereof will be considered as self-evident {\it a priori}, and it is
formally regarded as a physical device to incorporate the `classical'
basis into our formalism in a self-contained manner.
In this way we take account of 
a `macro-realistic' assumption that 
there exist in nature a special set of distinct states
whose linear superpositions is intrinsically prohibited.
This may be regarded as a postulate to be checked 
experimentally.
Within this framework, we shall next 
inquire
how small the effect of $H_1$ has to be, 
for the off-diagonal correlations to collapse spontaneously.

\section{Trigger problem}
\label{trigger}

The transformation (\ref{(1)}) is causal and reversible, while
the process (\ref{(2)}) is essentially irreversible,
in striking contrast.
With this crucial point in mind, 
we examine rather phenomenologically
under what condition the latter is triggered.
Taking up the problem this way, 
we are in an objective standpoint, 
regarding (\ref{(2)}) as a {\it spontaneous elementary process} 
inherent in Nature, which will occur
independently of 
any observer.
To characterize the process (\ref{(2)}),
we have to find relevant physical quantities.
It is naturally suspected that
the reduction process must have something to do with the Second Law of
Thermodynamics.\cite{rf:rp}

The first we can think of is the statistical entropy defined by
\begin{eqnarray}
 \Delta S&=&S_f-S_i
\nonumber\\
&=&
-\sum_n w_n \log w_n=
-{\rm Tr} (\hat{\rho}_f \log \hat{\rho}_f)
\end{eqnarray}
where $\hat{\rho}_f$ is the density matrix of the statistical {\it ensemble}
resulting from copies of
the initial state $\Psi$.
Here we used the notation $\Delta S$ to signify the change of 
entropy due to (\ref{(2)}), 
as we have $S_i=0$ for the initial pure states.
By the above expression, we still mean to represent
the entropy change in each {\it individual} event.
In effect, a single process (\ref{(2)}) transforms a pure state 
not into a mixed state, but to another pure state in general.
Still we associate $\Delta S$ not with the particular final state,
but with 
the final mixture of states, including those which
could have been but in fact not realized.
In short, by the entropy 
we characterize the {\it process},
not the {\it state}.
The entropy thus defined
characterizes the probabilistic process (\ref{(2)})
in an objective manner, 
and enables us to describe individual events in statistical terms.
It never decreases in (\ref{(2)}), 
while in (\ref{(1)}) holds $\Delta S=0$ identically.\cite{rf:jvn}
For this very reason, 
and since 
we obviously know that not all quantum states 
collapse spontaneously,
the appealing inequality $\Delta S>0$
borrowed from the Second Law 
is disqualified as the criterion for (\ref{(2)}).
Quite the contrary,
microscopic systems mostly preserve quite robust coherence
for good.
So we must seek another quantity.

The next to which we need pay due attention would be
the change of energy $\Delta E$,
defined similarly as $\Delta S$.
\begin{eqnarray}
 \Delta E&=& \sum_n w_n \langle\Phi_n|H|\Phi_n\rangle
-\langle\Psi|H|\Psi\rangle 
\nonumber\\
&=& -\sum_{m,n} c_m^*c_n\langle \Phi_m|H_1|\Phi_n\rangle
=-\langle \Psi|H_1|\Psi\rangle.
\label{DelE=}
\end{eqnarray}
This is the difference of statistical expectation values of energy;
the initial and final states, $\Psi$ and $\Phi_n$,
are generally not the eigenstates of the Hamiltonian $H$.
One may regard (\ref{DelE=}) as the coherence energy
shared by the initial linear combination,
but is lost in the final mixture.
It is the off-diagonal contribution of interaction energy
developed in (\ref{(1)}).
In particular, the reduction (\ref{(2)}) from the ground state of $H$
will always entail $\Delta E>0$.
The magnitude $|\Delta E|$ will characterize 
the strength of the collapsing coupling of 
distinct states in superposition.

Having thus discussed,
we had gone through 
a quintessential point intentionally tacitly.
That is,
by freely introducing $\Delta E$ for the reduction process (\ref{(2)}),
we are abandoning 
the topmost principle of physics, 
that is,
the law of conservation of energy. 
We claim that this is unpleasant but not unacceptable, 
since no fully accepted theoretical explanation has yet been given 
so far to the wave packet reduction.
In effect, we find no compelling reason, but inductive inference,
to conclude that
energy must conserve in (\ref{(2)}) as well.
Therefore,
in the following,
we shall dare to allow $\Delta E\ne 0$ as a working hypothesis, 
and discuss the notable consequences.

Now, we look for the condition 
in terms of $\Delta S$ {\it and} $\Delta E$.
Imagine a collection of 
a great number of reduction processes from a single definite state, 
regard them as physical and real processes, 
and try to make a thermodynamic description of them.
By natural inference, 
a criterion is drawn
on the analogy of the thermodynamic inequality of
irreversible processes in an {\it open system}, 
that is, 
\begin{equation}
\Delta S >\frac{\Delta E}{T_0},
\label{(4)}
\end{equation}
where $T_0$ is 
a constant with the dimension of energy.
To sum up, we hypothesize that
the wave packet reduction (\ref{(2)}) operates
when (\ref{(4)}) is met, or that,
{\it under the condition (\ref{(4)}), 
quantum states are ready to 
collapse 
spontaneously 
so as to provide the statistical ensembles
in conformity with the probability principle of QM.}
Accordingly, 
we interpret (\ref{(2)}) figuratively as depicting
an inherent tendency of quantum systems to behave
as if they were immersed in a heat bath of the temperature $T_0$.
One may then regard 
$-\Delta E/T_0$ as the entropy production in the heat bath, 
thereby the entropy principle recovered.
We claim $T_0$ is a {\it universal} constant.
In passing, it is of note that
the `thermodynamic' criterion (\ref{(4)}) is
fitted to accommodate
a holistic view on non-separability of quantum states.

As mentioned in the introduction, 
we are not concerned about the trigger mechanism of reduction.
It is presumed to be the universal process which sets in when the
above entropy criterion is met. 
To characterize how the stochastic process operates temporally,
we hypothesize that {\it reduction is an instantaneous Poisson process
with a mean frequency $\gamma_0$},
and we do not analyze it further.
Let this be 
contrasted with the GRW theory, in which
a reduction, especially for a macroscopic body,
is effected by 
numbers of instantaneous processes, called `hits'.
In each of the hits, wave function is multiplied by
a normalized Gaussian of width $\sim 10^{-5}$cm, 
and the hitting frequency $\lambda$ effectively depends on 
the number $N$ of constituent particles comprising the wave function,
$\lambda \simeq N\times 10^{-16}{\rm sec}^{-1}$.
In contrast, besides $T_0$,
we regard $\gamma_0$ as
another universal constant, independent of the system size.\footnote{
One may argue that the frequency $\gamma_0$
of reduction and the energy scale $T_0$
that we introduced above 
must be related
by an uncertainty relation, $T_0\sim \hbar \gamma_0$.
However, the argument would be presumably based on 
the very assumption we relinquished at the outset, 
that the reduction is cognate with
the Schr\"odinger-like time evolution.
Therefore, the presumed relation is
no more than an unfounded hope
from our standpoint to accept the two types of processes as 
essentially distinct and equally elementary.
}
The size dependence will be manifested
through the criterion (\ref{(4)}).
By giving up inquiries about the reduction mechanism, or
by the definition of `classical' states,
we get free from the problem of the tails of the reduced wave
function as was raised against the GRW theory.~\cite{rf:bg}

\section{Discussion}
\label{discussion}

In the above formulation,
the unitary evolution is interrupted by 
a reduction within a finite interval of time
$\tau_0 =\gamma_0^{-1}$ after the criterion is met.
Hence
we predict results in disaccord with 
those without reduction 
after the elapse of time
$t>\tau_0$,
where we can no longer expect the interference phenomena 
due to the off-diagonal correlation inscribed in (\ref{DelE=}),
{\it not for all practical purposes but in principle}.
Therefore, we may say that
$\tau_0$ is a characteristic time scale
separating the coherent reversible quantum regime $t<\tau_0$
from the incoherent irreversible `classical' regime. 
In the weak coupling limit of $H_1$,
time evolution in the latter regime will be described by 
infrequent {discontinuous} 
jumps between diagonal states of $H_0$,
or by the Pauli master equation (Appendix \ref{mastereq}).
The incoherent regime appears only when 
one can conceive of such processes allowed by the criterion (\ref{(4)}).

In this regard,
it is remarked that the entropy criterion
(\ref{(4)}) is always met for the energy conserving collapse, 
namely, for $\Delta E=0$.
This may be the case in which the Fermi golden rule (\ref{Wjk})
applies,
where we claim that
{\it the generalized entropy principle applies to
microscopic processes as well}.
This constitutes a part of 
non-trivial contents of our proposition.
The entropy principle decides whether or not
a classically inconceivable linear combination 
is projected onto a classically interpretable 
constituent.
This interpretation seems more convincing than the conventional
one that a body cannot be in a definite macro-state until it is
observed by an observer.\footnote{
To preach the apparent link between
the entropy principle and the measurement,
one might respond as follows; 
it is obvious that when the process creates entropy according
to the observer, an irreversible process has taken place (according to
the observer).
But then, he must contemplate
how the entropy is created, or 
from where
on earth 
emerges {\it the} irreversibility.
Our point of view is that {\it all} the irreversible processes known to
date, microscopic or macroscopic, are 
governed by the entropy principle.
This is no way obvious.
It is underlined that
we regard `observation', or
`measurement', as secondary to increase of entropy,
not the other way around.
In our hypothetical formulation,
we introduce a dualism in the laws of quantum mechanics,
but we make it up for a unified description of microscopic and
macroscopic irreversibility.
}

In general,
we may have a mixed state for
the initial state of (\ref{(2)}).
If we have initially 
the canonical ensemble at temperature $T$,
it is found that (\ref{(4)}) is not met unless 
$T$ is lower than $T_0$, or $T\alt T_0$.
In fact,
as shown below in (\ref{2cond}),
the effects of undesirable energy non-conservation 
come to the fore only 
in the low energy section $T$, $\Delta E\alt T_0$.
To put it definitely,
the ground state comprising weakly coupled parts,
or the stationary state with $\Delta E\alt T_0$, 
becomes intrinsically unstable
with the lifetime $\tau_0$,
while strongly bound states which necessarily meet $\Delta E\gg T_0$
will remain intact practically.

This is the gist of how
we reconcile QM with macro-objectivism,
i.e., 
by destabilizing a superposition state of Schr\"odinger's cat.
Quantum correlation will not proliferate without limit.
A sharp distinction between stable and unstable
is drawn in comparison with the energy scale $T_0$.
The consequences as discussed above strongly suggest that
$T_0$ must be 
substantially lower than typical microscopic energy scales
and $\tau_0$ be larger than any of microscopic time scales.
In principle, these constants have to be determined experimentally.
Some examples for this purpose are discussed below.

Let us consider an isolated two-state system
described effectively by the Hamiltonian
\begin{equation}
 H_1=-v_1
\left(
|\uparrow\rangle\langle\downarrow|
+|\downarrow\rangle\langle\uparrow|
\right),
\end{equation}
where $v_1>0$, and let $|\uparrow \rangle$ and $|\downarrow \rangle$ 
define the `classical' basis.
The ground state 
is given by the linear superposition,
\[
|0\rangle=
\frac{1}{\sqrt{2}}
\left(|\uparrow\rangle+|\downarrow\rangle \right).
\]
In equilibrium, the instability condition (\ref{(4)}) gives
\begin{eqnarray}
&T_0 \log 2> v_1,\qquad  &v_1 \gg T
\nonumber\\
&{T_0} > 2T.\qquad   &T\gg v_1
\label{2cond}
\end{eqnarray}
Under the condition,
we can derive a dynamical equation for the density matrix
$\rho_{\sigma\sigma'}$.
In terms of ${\omega_0}=v_1/\hbar$,
for 
$w\equiv \rho_{\uparrow\uparrow}-\rho_{\downarrow\downarrow}$ 
we obtain
\begin{equation}
 \ddot{w}+\gamma_0 \dot{w}+
\omega_0^2
w=0,
\end{equation}
while 
for $\delta\rho_{\pm}=\rho_{\uparrow\downarrow}\pm\rho_{\downarrow\uparrow}$,
\begin{eqnarray}
 \dot{\delta\rho_-}&=&
-i\omega_0
w-\gamma_0\delta\rho_-,
\\
 \dot{\delta\rho_+}&=&
-\gamma_0\delta\rho_+.
\end{eqnarray}
By the decay rate $\gamma_0$, the system tends to 
the diagonal equilibrium $w=\delta \rho_\pm=0$.
In general, the criterion (\ref{(4)}) is expressed in terms of 
$\Delta E=v_1\delta \rho_+$ and
$\Delta S$, a function of $w$ and $\delta \rho_\pm$.
In the equilibrium, 
one would rather follow
the state vector as a function of time.
In the `classical' limit $\omega_0\tau_0\ll 1$,
there appears an intermediate time scale $\Delta t$
between $\tau_0$ and $\omega_0^{-1}$;
 $\tau_0\ll \Delta t\ll \omega_0^{-1}$.
Once localized in $|\uparrow \rangle$ or $|\downarrow\rangle$,
then the gross observers who can tolerate 
an inaccuracy of $\Delta t\agt \tau_0$
will find infrequent discontinuous random switching
back and forth
between  the classical states
$|\uparrow \rangle$ and $|\downarrow\rangle$,
fluctuating with the elongated time scale $\omega_0^{-2}\tau_0^{-1}$.

Let us speculate further what would occur if 
the system 
under the condition (\ref{2cond})
is open to surrounding infinite space via some interaction,
i.e., when the system cannot reach statistical equilibrium.
Suppose an interaction $H_2$ to relax
the excited state
$|1\rangle=\left(|\uparrow\rangle-|\downarrow\rangle \right)/\sqrt{2}$
back to $|0\rangle$
within a `microscopic' lifetime $\tau$,
presumably by emitting a `photon' $|\omega\rangle$, e.g.,
\begin{equation}
H_2= v_2 \left(
|\omega\rangle|0\rangle\langle 1|+
|1\rangle\langle 0| \langle \omega|
\right).
\end{equation}
In the limit $\tau  \ll \tau_0$,
reduction is relatively insignificant,
and the ground state
$|0\rangle$, even if unstable, survives 
for a while.
After the elapse of time $\tau_0$, 
it is spontaneously excited to populate
$|1\rangle$, which then immediately
relax back again to $|0\rangle$ within the lifetime $\tau$.
Hence, 
under the condition (\ref{2cond}),
we predict 
infinite cycles of reduction and causal evolution.
The cyclic time evolution,
adapted to the wave propagation in free space,
has been noted in the K\'arolyh\'azy model.\cite{rf:qg}
In the above example,
not only an explicit demonstration of energy non-conservation,
but also the proposed statistics of reduction
with the mean lifetime $\tau_0$,
are reflected 
in the counting of the emitted state $|\omega\rangle$
{\it in principle.}

In practice,
a macroscopic body must always have a 
well defined position 
in the objective description of reality.
In our scheme, a spatially localized quasi-eigenstate of a
massive particle in free space
is realized similarly as above;
in this case,
at the cost of 
the last off-diagonal 
correlation
left in the Hamiltonian, namely, 
the kinetic energy
\begin{equation}
H_1=-\frac{\hbar^2}{2m}\left(
\frac{\partial^2}{\partial x^2}
+\frac{\partial^2}{\partial y^2}
+\frac{\partial^2}{\partial z^2}
\right),
\end{equation}
for which 
the criterion (\ref{(4)}) reads $v \agt \lambda_0^3$,
where 
$v$ is the volume occupied by the wave packet
and 
\begin{equation}
 \lambda_0=\frac{h}{\sqrt{2\pi m T_0}}
\end{equation}
is the thermal de Broglie wavelength at the temperature $T_0$.
The reduced wave packet will unitarily develop 
further for the duration $\tau_0$ at most,
or it will grow as large as 
$\lambda_{0,{\rm max}}$ given by
\begin{equation}
\lambda_{0,{\rm max}}^2=
{\lambda_0^2 +
\left(
\frac{\hbar \tau_0}{2m\lambda_0}\right)^2
}.
\end{equation}
The stochastic processes 
keep the wave packet from spreading without limit.
Note
$\lambda_{0,{\rm max}}\simeq \lambda_0$ if $T_0\tau_0\ll 4\pi
\hbar$ (cf. the second footnote).

The above consideration is directly applied to 
the center-of-mass motion of the many-body wave function of a
macroscopic body.
The body will perform a Brownian motion within the width of
order $\lambda_0$ along the classical trajectory,
as it is expected from Ehrenfest's theorem.
As a matter of fact, 
the results of the last paragraph remain valid
as far as the potential energy $H_0=V(x,y,z)$
does not vary appreciably 
over a region of the linear dimension $\sim \lambda_0$.
Thus, our predictions are qualitatively similar as 
in the GRW model
and the K\'arolyh\'azy model,\cite{rf:grw,rf:qg}
though quantitatively the results for the width are all different from
each other.\cite{rf:af}
For macroscopic bodies,
the finite breadth $\lambda_0$ of the trajectory
means slight departure from classical mechanics.\cite{rf:grw,rf:qg,rf:af}
For microscopic systems,
the width $\lambda_0$ of a wave packet 
must be reflected in diffraction experiments
as a washout of an interference pattern,
although we predict no phase shift as expected for nonlinear
theories.\cite{rf:as}
The effect of $\lambda_0$ will be manifested as
the mass dependence on 
the contrast of interference fringes for material waves of a fixed de
Broglie wavelength $\lambda>\lambda_0$.

The upper bound allowed for the numerical choice of $T_0$
is set by experiments on quantum interference.
A double slit neutron interference experiment 
by Zeilinger $et$ $al.$,\cite{rf:az}
in quantitatively agreement with the
prediction of QM,
suggests $\lambda_0 > 243 {\rm \AA}$,
a coherence length of the diffracted neutrons,
or $T_0< 7\times 10^{-26}{\rm J}$.
In practice, the bound should be 
still much lowered considerably,
by reducing the bandwidth $\Delta \lambda =1.4 {\rm \AA}$ of the
neutron.
%
%
%
%
%
On the other hand,
the lower bound of $T_0$ is set from a rough estimate 
on a macroscopic body, 
e.g.,
$\lambda_0\alt 1\mu$m for $m=1$ng, 
by which $T_0\agt 10^{-49}$J.
Therefore, 
as anticipated by
the fundamental `shiftiness' of the micro-macro boundary
in the standard quantum theory,
we still have a wide range left unexplored for
$T_0$.
This holds true even if a free neutron is
supposed to have $\lambda_0>1$m.

As for the time scale $\gamma_0=\tau_0^{-1}$, 
we claim that a non-trivial aspect of our proposition $\gamma_0>0$
has been already supported 
experimentally by an exponential decay law of
an unstable
system like a radioactive nucleus.
Indeed, as is well known,
the decay law $N(t)=N(0) \exp(-t/\tau)$
for the number $N(t)$ of radioactive nuclei 
cannot derive strictly from the unitary time evolution
of the decaying state.\cite{rf:fgr}
Instead, it follows from the classical assumption of the 
complete independence of 
the nuclei state of the past history, $N(t+t')=N(t)N(t')$.
Therefore, we agree with
Fonda $et$ $al.$\cite{rf:fgr} and others, who point out that
the exponential behavior is explained satisfactorily 
by random reduction processes due to `measurement',
although we disagree with them on the crucial point that 
the frequent `measurement' processes 
be ascribed to interactions with
its environment.
On the contrary,
one would rather expect 
the lifetime $\tau$ of the unstable nucleus
is an intrinsic property of the nucleus, 
independent of the presence or absence of any `observer'.

Based upon common sense, 
we suspect that 
$\tau_0$ would be at large of order of the human perception time,
or $\tau_0\alt 10^{-2}$ second.
Furthermore, 
we have to remark 
that we are severely put under the
constraint that the power of anomalous energy fluctuation
$\sim T_0/\tau_0$,
that we predict rather unwillingly,
must be extraordinarily small
for the present theory to be viable.

\section{Summary}
\label{summary}

To summarize, we postulate
two fundamental laws of time evolution in quantum mechanics.
The first is causal and unitary, described by 
the Schr\"odinger equation.
The second
is unprecedented as one of 
the elementary processes; irreversible, stochastic,
and energy non-conserving.
It is proposed that
quantum state $\Psi$ of the Hamiltonian $H=H_0+H_1$ is
spontaneously and instantaneously
projected onto one of the eigenstates $\Phi_n$ of
the reduced Hamiltonian $H_0$
with the probability $|\langle \Phi_n |\Psi \rangle|^2$
and with the mean frequency $\gamma_0$,
when the generalized entropy criterion (\ref{(4)}) is satisfied.

\section*{Acknowledgement}

It is a pleasure to express my gratitude to 
Professor A.~J.~Leggett 
whose comments on the first draft, particularly on the preferred basis problem,
and the encouragement above all are 
invaluable to complete this work.
I also wish to thank Prof. B.~d'Espagnat for 
a comment and a pointer for dynamical reduction theories.

\appendix
\section{master equation}
\label{mastereq}

In this appendix,
we derive a master equation for 
the density matrix $\hat{\rho}(t)$
on the basis of the formalism presented in this paper.
We assume the representation in which $H_0$ is diagonal
while $H_1$ has no diagonal element,
and the latter is treated by perturbation theory.
The derivation is then compared with
the technique due to Van Hove.\cite{rf:lvh}

To show the non-trivial effect of the constant $\gamma_0$,
we consider a special case in which 
the generalized entropy criterion (\ref{(4)})
is always met for 
the small perturbation $H_1$.
Then, by prescription,
the effect of reduction is described by 
the equation
\begin{equation}
 \frac{\partial \hat{\rho}(t)}{\partial t} =\frac{1}{i\hbar} 
[H,\hat{\rho}]-\gamma_0 \delta \hat{\rho},
\label{eqofmo}
\end{equation}
where
\begin{equation}
\delta \hat{\rho}
(t)= \hat{\rho}(t)-\sum_n P_n \hat{\rho}(t)P_n,
\end{equation}
and 
$P_n=|n\rangle\langle n|$
is the projection operator on the $n$-th eigenstate
$|n\rangle$ of $H_0$.
The last term of (\ref{eqofmo})
effectively suppresses the off-diagonal elements of $\hat{\rho}$.
Eq.~(\ref{eqofmo}) was specifically discussed 
by Fonda, Ghirardi and Rimini\cite{rf:fgr0}
to investigate the environmental effect of 
random `measurements'
on the decay of an unstable state.
On the basis of their results, we only have to investigate
time evolution from general states,
instead of a special initial state.
Nevertheless, it is stressed that
we interpret (\ref{eqofmo}) 
quite differently from Fonda $et$ $al$.
The last term of (\ref{eqofmo})
is effective only under a prescribed condition,
and we regard it as an intrinsic,
not extrinsic, property of the system.

Let us consider the diagonal elements of 
\(
 \Delta \hat{\rho}
= \hat{\rho}(\Delta t)-\hat{\rho}(0),
\)
for which (\ref{eqofmo}) is solved by iteration.\cite{rf:fgr0}
\begin{eqnarray}
 \Delta \rho_{jj}
&=& 
\frac{2}{\hbar^2}
\sum_k \frac{|\langle j|H_1| k\rangle|^2}{\omega_{jk}^2+\gamma_0^2}
\nonumber\\
&&\times \biggl[
\gamma_0 \Delta t
+\frac{\omega_{jk}^2-\gamma_0^2}{\omega_{jk}^2+\gamma_0^2}
\left(
1-e^{-\gamma_0 t}\cos(\omega_{jk}\Delta t)\right)
\nonumber\\
&&-\frac{2\gamma_0\omega_{jk}}{\omega_{jk}^2+\gamma_0^2}
e^{-\gamma_0 t}
\sin(\omega_{jk}\Delta t)
\biggr]
\left(\rho_{kk}-\rho_{jj}\right),
\label{Delrhojj}
\end{eqnarray}
where $\hbar\omega_{jk}=E_j-E_k$ and 
$E_j$ is the unperturbed energy of $|j\rangle$.
The diagonal elements of $\hat{\rho}$ are stationary
in the zeroth order approximation.
In the long time regime $\Delta t\gg \tau_0=\gamma_0^{-1}$,
from (\ref{Delrhojj}),
we immediately obtain the master equation
\begin{equation}
 \frac{\Delta \rho_{jj}}{\Delta t}
=\sum_k 
\left(W_{jk}\rho_{kk}-W_{kj}\rho_{jj}\right),
\label{master}
\end{equation}
where we define the transition probability
\begin{equation}
 W_{jk}=
\frac{2\pi}{\hbar}
|\langle j|H_1| k\rangle|^2
\delta_{\gamma_0}(\hbar\omega_{jk}),
\label{Wjk}
\end{equation}
in terms of
\begin{equation}
\delta_{\gamma_0}(\omega)
=\frac{\gamma_0}{\pi(\omega^2+\gamma_0^2)}.
\label{delgam0}
\end{equation}
It is remarked that
the effect of $\gamma_0$ on 
the off-diagonal matrix elements
is to replace
the time dependent factor $e^{-i\omega_{ij}t}$
with $e^{-i\omega_{ij}t}e^{-\gamma_0t}$,
so that they are suppressed effectively.

Next, 
let us set $\gamma_0=0$ in (\ref{Delrhojj})
to discuss a derivation without $\gamma_0$.
To derive the master equation,
Van Hove assumes the peculiar weak coupling limit
$\langle H_1\rangle\rightarrow 0$ while $\Delta t\rightarrow \infty$, 
so as to fix
$|\langle j|H_1| k\rangle|^2 \Delta t$ as constant.\cite{rf:lvh}
Indeed, in this limit too,
one recovers the same equation (\ref{master}) formally,
but with the genuine delta function $\delta(\omega)$,
owing to the formula
\[
 \lim_{ t\rightarrow \infty} \frac{1-\cos \omega t}{t\omega^2}
=\pi \delta(\omega),
\]
instead of the above Lorentzian $\delta_{\gamma_0}(\omega)$
for $W_{jk}$.
But this is only the first half of the derivation.

In order to warrant the absence of interference terms,
Van Hove has to restrict the initial configuration for $\hat{\rho}(0)$
to a special class of states.
In effect, such a special assumption 
on the absence of the initial correlations,
or on the initial states of low entropy,  
is surely imperative to derive the irreversible equation 
on the basis of reversible mechanics,
for it is obviously generally impossible.

Incidentally,
the indeterminacy of energy $\sim \hbar\gamma_0$,
as implied in (\ref{delgam0})
for individual processes, is 
completely within the conventional framework of quantum mechanics.  
This effect should not be misidentified with 
(statistically significant) energy non-conservation $\Delta E\ne 0$ 
that we invoke for (\ref{DelE=}).

\end{document}